\documentclass[review]{elsarticle}

\usepackage{lineno,hyperref}
\usepackage{color}

\modulolinenumbers[5]

\newcommand{\subf}[2]{%
  {\small\begin{tabular}[t]{@{}c@{}}
  #1\\#2
  \end{tabular}}%
}

\biboptions{sort&compress}
\begin{document}

\begin{frontmatter}

\title{Temporal code versus rate code \\ for binary Information Sources}

\author[mymainaddress]{Agnieszka Pregowska}

\author[mymainaddress]{Janusz Szczepanski \corref{mycorrespondingauthor}}
\cortext[mycorrespondingauthor]{Corresponding author}
\ead{jszczepa@ippt.pan.pl}

\author[mymainaddress]{Eligiusz Wajnryb}

\address[mymainaddress]{Institute of Fundamental Technological Research, Polish Academy of Sciences, Pawinskiego 5B, Warsaw}

\begin{abstract}
Neuroscientists formulate very different hypotheses about the nature of neural code. At one extreme, it has been argued that neurons encode information in relatively slow changes of individual spikes arriving rates (rates codes) and the irregularity in the spike trains reflects noise in the system, while in the other extreme this irregularity is the code itself (temporal codes) thus the precise timing of every spike carries additional information about the input. It is well known that in the estimation of Shannon information transmission rate the patterns and temporal structures are taken into account, while the "rate code" is already determined by the firing rate, i.e. by spike frequency. In this paper we compare these two types of codes for binary Information Sources which model encoded spike-trains. Assuming that the information transmitted by a neuron is governed by uncorrelated stochastic process or by process with a memory we compare the information transmission rates carried by such spike-trains with their firing rates. We showed that the crucial role in studying the relation between information and firing rates is played by a quantity which we call "jumping" parameter. It corresponds to the probabilities of transitions from no-spike-state to the spike-state and vice versa. For low values of jumping parameter the quotient of information and firing rates is monotonically decreasing function of firing rate, thus there is straightforward, one-to-one, relation between temporal and rate codes. On the contrary, it turns out that for large enough jumping parameter this quotient is non-monotonic function of firing rate and it exhibits a global maximum, i.e. in this case there exists the optimal firing rate. Moreover, there is no one-to-one relation between information and firing rates, so the temporal and rate codes differ qualitatively. This leads to the observation that the behavior of the quotient of information and firing rates for large jumping parameter is especially important in the context of bursting phenomena.
\end{abstract}


\end{frontmatter}


\section{Introduction}
Fundamental Neuroscience problem is to understand how neurons encode and process information \cite{Stein_1965,Rieke_1997, van_Hemmen_Sejnowski_2006}. In general it is not easy to determine the neural code structure. Since Adrian's experiments \cite{Adrian_1926} which established that individual sensory neurons produce action potentials, or spikes it is assumed that a single neuron provides information just through spikes sequence, i.e. spike-trains. Although it is now generally accepted that a spike sequence is the way the information is coded by a single neuron, the structure and the mechanisms of code formation are still mysteries. In 1976 Burns and Webb \cite{Burns_Webb_1976} for the first time showed that the total number of emitted spikes arrives in a highly irregular manner. When the same stimulus is applied repeatedly the number of spikes varies substantially from trial to trial \cite{Richmond_1990}. This has inclined Neuroscientists to formulate very different hypotheses about the nature of the neural code. Two main ideas, not excluding each other, are of special interest. The first theory is based on the idea of "temporal code" \cite{Coop_Reeke_2001,van_Hemmen_Sejnowski_2006, Duguid_2006, Yu_2014} and goes into the spike -trains structure while the second referred to as "rate code" theory \cite{Stein_1965, Ricciardi_1979, Lansky_2001, van_Hemmen_Sejnowski_2006,  Di_Maio_2008} assumes that the neural code is embedded in the spike frequency, defined as the number of spikes emitted per second. The temporal coding mechanism, which builds a relationship of temporal process between the output firing patterns and the inputs of the nervous system, has received much attention \cite{Gerstner_1997,Goldberg_Andreou_2004, Butts_2007}. On the other hand in the transfer  of information the most expansive energetically is spiking process  \cite{Ames_2000, Moujahid_d_Anjou_2012}, thus in the first approximation the firing rate can be treated as the energy marker. Inspired by the thermodynamics \cite{Van_Kampen_2007} we also consider the derivative of entropy over energy which is the analog of temperature inverse.
\par
In this article we give the theoretical insight into understanding the neural code nature, namely we study this problem for two types of binary Information Sources. Assuming that the information transmitted by a neuron is governed by uncorrelated stochastic process or by process with a memory we study the relation between the Information Transmission Rates ITR carried by such spike-trains and their firing rate $F_{R}$. To this end the Information-Firing-Quotient $IFQ$, being the ratio of information and firing rate, is introduced in Section \ref{info}. For large $IFQ$ transmission is more optimal in the sense of information amount transmitted at the cost of unit energy. We show that the crucial role in studying $IFQ$ properties is played by the "jumping" parameter. This parameter is the sum of transition probabilities from no-spike-state to spike-state and vice versa. We show that for the low values of jumping parameter the quotient of information and firing rates is monotonically decreasing function of firing rate, thus there is straightforward, one-to-one, relation between temporal and rate codes. On the contrary, it turns out that for large enough jumping parameter this quotient is non-monotonic function of firing rate and it exhibits well pronounced global maximum. Thus, in this case the optimal firing rate exists. Moreover, there is no one-to-one relation between information and firing rate and the temporal and rate codes differ qualitatively. The behavior of the quotient of information and firing rates for large jumping parameter is especially important in the context of bursting phenomenon\cite{Mukherjee_1995, Reinagel_1999, Gerstner_2014}.
\par
The paper is organized as follows. In Section \ref{info} the basic concepts of Information Theory and formulas concerning Bernoulli and Markov processes are briefly recalled. The comparison of information transmission and firing rates for these processes is presented in Section \ref{results}. The last Section contains discussion and conclusions.
\section{Information Theory in Neuroscience}\label{info}
In Neuroscience the information transfer is quantified by many authors in terms of Information Theory \cite{Borst_1999, van_Hemmen_Sejnowski_2006}. In general, neuronal communication systems are represented by Information Source, communication channel and output signals \cite{Shannon_Weaver_1963, Ash_1965, Cover_Thomas_1991}. Both messages coming from Information Source and output signals are represented by sequences of symbols \cite{Bialek_1991, Strong_1998, Borst_1999, van_Hemmen_Sejnowski_2006, Gerstner_2014}. These sequences can be understood as trajectories of stationary stochastic processes. In this paper we study the Information Sources which are represented by Bernoulli or Markov processes \cite{Feller_1958,Bialek_1991}.
\paragraph{Entropy}
First, we briefly recall the fundamental concepts of Information Theory \cite{Ash_1965, Shannon_Weaver_1963, Cover_Thomas_1991} that are adapted to Neuroscience issues. Let $Z^{L}$  be a set of all words (i.e. blocks) of length $L$, built of symbols (letters) from some finite alphabet $Z$. Each word $z^{L}$ can be treated as a message sent by Information Source ${Z}$ being a stationary stochastic process. If $P(z^{L})$ denotes the probability the word $z^{L}\in Z^{L}$ occurs, then the information in Shannon sense carried by this word is defined as
\begin{equation}\label{ITR}
I(z^{L}):=-\ln{P(z^{L})} \ . 
\end{equation}
In this sense, less probable events carry more information. We use the natural logarithm to get more compact form of the formulas. In case when logarithm to the base 2 is used, the factor of $\ln{2}$ appears. Expected or average information of $Z^{L}$, called Shannon block entropy reads
\begin{equation}\label{entropy}
H(Z^{L}):=-\sum\limits_{z^{L}\in Z^{L}}P(z^{L})\ln{P(z^{L})} \ . 
\end{equation}
Since the word length $L$ can be chosen arbitrary, the block entropy does not perfectly describe the Information Source \cite{Ash_1965, Cover_Thomas_1991}. The more adequate characteristics of the information transmission rate is defined in the next Subsection. For the special case of two-letter alphabet $Z=\{0,1\}$ and the length of words $L=1$ we introduce the following notation for the entropy
\begin{equation}\label{H1}
H_{1}(p):=H(Z^{1})=-p\ln{p}-(1-p)\ln{(1-p)} \ ,
\end{equation}
where $P(1)=p, P(0)=1-p$ are the associated probabilities.
This is, in fact, formula for the entropy of two-state system. 
\paragraph{Information Transmission Rate}
The entropy of spike trains themselves evaluates how much information these spikes could provide. The adequate measure for estimation of efficiency of information source is the information transmitted in average by a single symbol. This measure, which characterizes Information Source $\{Z\}$, is called Information Transmission Rate $ITR$ and is defined as \cite{Ash_1965, Cover_Thomas_1991}
\begin{equation}\label{ITR_3}
ITR(\left \{Z \} \right):=\lim_{L \to \infty} \frac{H(Z^{L})}{L} \ .
\end{equation}
Information transmission rate exists if and only if the stochastic process $\{Z\}$ is stationary \cite{Cover_Thomas_1991}.
\par
The Information Transmission Rate is very important quantity especially due to the Asymptotic Equipartition Theorem. This theorem states that information per symbol for most of the messages coming from a given source is close to $ITR$ \cite{Ash_1965, Cover_Thomas_1991}.
\paragraph{Firing Rate}
Since the experiment of Adrian \cite{Adrian_1926} very important characterization of both neural network dynamics and neural computation is the firing rate $F_{R}$ of spike-trains. The first and most commonly used definition of the firing rate refers to temporal average \cite{Gerstner_1997, van_Hemmen_Sejnowski_2006, Gerstner_2014} and reads
\begin{equation}\label{fr}
F_{R}=\frac{n_{T}}{T} \ ,
\end{equation}
where $n_{T}$  denotes spike count and $T$ is time window length. In practice, in order to get sensible averages, some reasonable number of spikes should occur within the time window. Since the messages are trajectories of stationary stochastic process the firing rate as defined by (\ref{fr}) is specyfic for a given Information Source provided $T$ is large enough. Thus, $F_{R} \cdot \Delta \tau$ can be identified with the probability p of spike appearance, where $\Delta \tau$ is the time resolution or bin size. 
\paragraph{Bernoulli process}
Assuming the size of bin spike trains can be encoded \cite{Bialek_1991} in such a way that 1 is generated with probability $p$ (spike is arrived in the bin), 0 is generated with probability $1-p$ (spike is not arrived). In the situation when consecutive bits in message are uncorrelated, we are in the regime of Bernoulli process \cite{Feller_1958, Cover_Thomas_1991}. Following the entropy definition (\ref{entropy}) the Information Transmission Rate (\ref{fr}) for Bernoulli process reads 
\begin{equation}\label{ITR_2}
ITR(p)=-p\ln{p}-(1-p)\ln{(1-p)}=H_{1}(p) \ .
\end{equation}
Further in the paper the Bernoulli process will be considered as a benchmark for more complex processes like the Markov ones.
\paragraph{Markov process}\label{markov_process}
Again assuming the size of bin we now consider as the Information Source the discrete-time, two-state Markov processes. The conditional probabilities for such processes are completely defined in terms of two transition probabilities from state 0 to state 1, $p_{1|0}$ and from state 1 to state 0, $p_{0|1}$. The Markov transition  probability matrix $P$ can by written as
\begin{equation}\label{markov}
P:=
\left[
\begin{array}{ccc}
1-p_{1|0} & p_{0|1}
\\
&
\\ 
p_{1|0} & 1-p_{0|1}
\nonumber
\end{array}
 \right] \ .    
\end{equation}
We assume here that the process is homogeneous in time.
\par
The probability evolution is governed by Master Equation \cite{Van_Kampen_2007}
\begin{equation}\label{markov2}
\left[\begin{array}{ccc}
P_{n+1}(0) \\ & \\ P_{n+1}(1)
\end{array} \right]=
\left[\begin{array}{ccc}
1-p_{1|0} & p_{0|1}
\\
&
\\ 
p_{1|0} & 1-p_{0|1}
\nonumber
\end{array}
 \right]\           
\cdot
\left[\begin{array}{ccc}
P_{n}(0) \\ & \\ P_{n}(1)
\end{array} \right]\ , 
\end{equation}       
where $n$ stands for the discrete time, and the stationary solution reads
\begin{equation}
\left[\begin{array}{ccc}
P_{eq}(0) \\ & \\ P_{eq}(1)
\end{array} \right]=
\left[\begin{array}{ccc}
p_{0|1}/(p_{0|1}+p_{1|0})
\\ & \\
p_{1|0}/(p_{0|1}+p_{1|0})
\end{array} \right]\ .
\end{equation}
The Information Transmission Rate (\ref{ITR_3}) of such a Markov source reads \cite{Cover_Thomas_1991}.
\begin{eqnarray}
ITR=P_{eq}(0)(-p_{1|0}\ln{p_{1|0}}-(1-p_{1|0})\ln{(1-p_{1|0})}) \nonumber
\\
+P_{eq}(1)(-p_{0|1}\ln{p_{0|1}}-(1-p_{0|1})\ln(1-p_{0|1}))\ ,
\end{eqnarray}
or making use of notation (\ref{H1}), in compact form
\begin{equation}\label{itr_markov_case}
ITR=P_{eq}(0)H_{1}(p_{1|0})+P_{eq}(1)H_{1}(p_{0|1}) \ .
\end{equation}
For the later use the probability of state "1" is for short denoted by $p$
\begin{equation}\label{p_markov_case}
p:=P_{eq}(1)=\frac{p_{1|0}}{(p_{0|1}+p_{1|0})} \ .
\end{equation}
and in fact is interpreted as the firing rate.
Please, note that for the special case when $p_{0|1}+p_{1|0}=1$ the Markov process becomes uncorrelated and reduces to Bernoulli process with $p=p_{1|0}$.
\section{Results}\label{results}

It is well known that the "temporal code" approach requires the reliable estimation of Information Transmission Rate which must take into account the patterns and temporal structures  \cite{Strong_1998, Kontoyiannis_1998, Amigo_2004, Arnold_2013, Lempel_Ziv_1976_a, Amigo_2015, Pregowska_2015}, while the "rate code" is determined just by the firing rate, which in turn is fully given by the probability $p$. Addressing the problem of relation between "temporal code" and "rate code" we introduce the Information-Firing-Quotient $IFQ$ defined as the ratio
\begin{equation}
IFQ:=\frac{ITR}{F_{R}} \ .
\end{equation}
$IFQ$ can be understood as the information cost in terms of the energy units. Further we analyze the $IFQ$ for messages coming from two qualitatively different Information Sources, namely Bernoulli and Markov processes. 
\begin{figure}[h!]\label{fig1}
     \begin{tabular}{|c|c|}
\hline
\multicolumn{1}{|l|}{a)} & \multicolumn{1}{|l|}{b)}
\\
\subf{\includegraphics[width=0.5\textwidth]{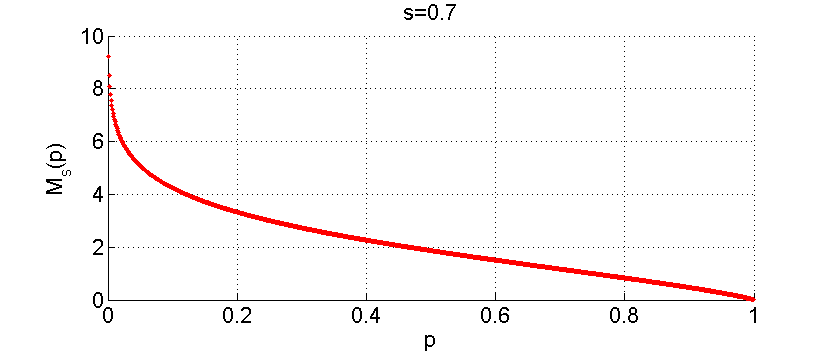}}{}&
\subf{\includegraphics[width=0.5\textwidth]{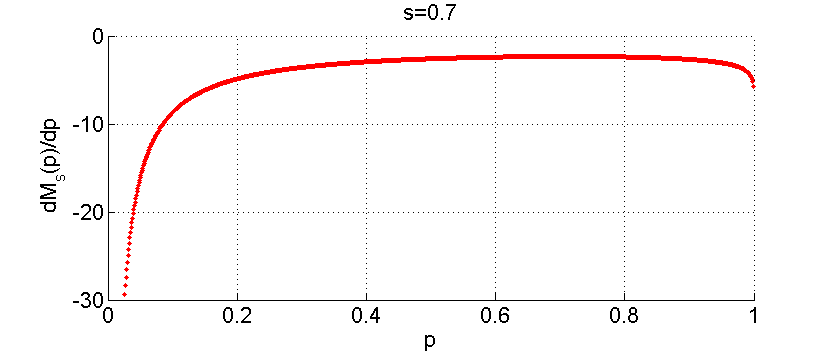}}{}
\\
\hline
\multicolumn{1}{|l|}{c)} & \multicolumn{1}{|l|}{d)}
\\
\subf{\includegraphics[width=0.5\textwidth]{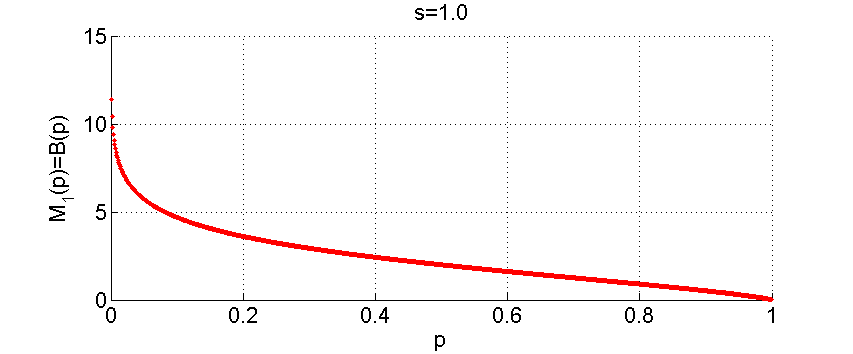}}{}&
\subf{\includegraphics[width=0.5\textwidth]{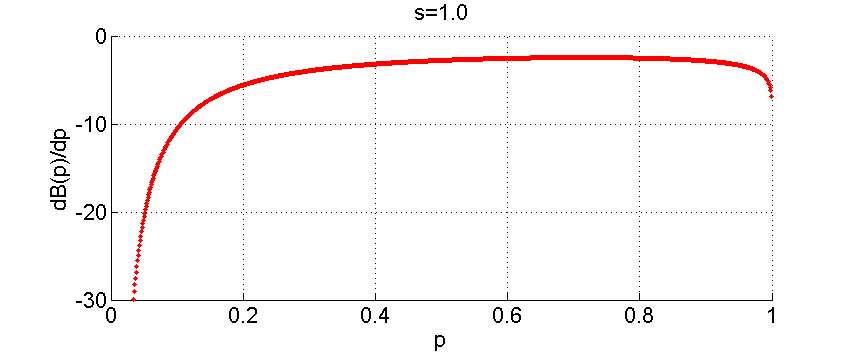}}{}
\\
\hline
        \end{tabular}
  \caption{The typical course of Information-Firing-Quotient $M_{s}$ versus spiking probability $p$ for low values of jumping parameter $s \le 1$. In panels a, b course of $M_{s}$ and its derivative for $s=0.7$ is shown. Panels c, d the course of $M_{s}$ and its derivative for $s=1.0$ is presented. Notice that $s=1$ is the Bernoulli source case.
}\label{fig1}
      \end{figure}
\begin{figure}[h!]\label{fig2}
     \begin{tabular}{|c|c|} 
\hline
\multicolumn{1}{|l|}{a)} & \multicolumn{1}{|l|}{b)}
\\
\subf{\includegraphics[width=0.5\textwidth]{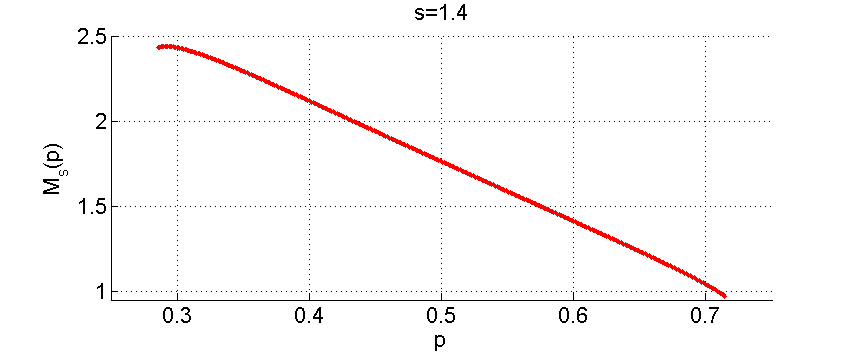}}{}&
\subf{\includegraphics[width=0.5\textwidth]{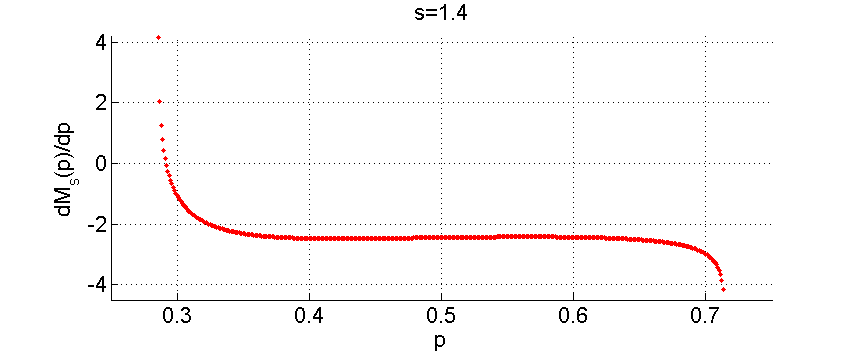}}{}
\\
\hline
\multicolumn{1}{|l|}{c)} & \multicolumn{1}{|l|}{d)}
\\
\subf{\includegraphics[width=0.5\textwidth]{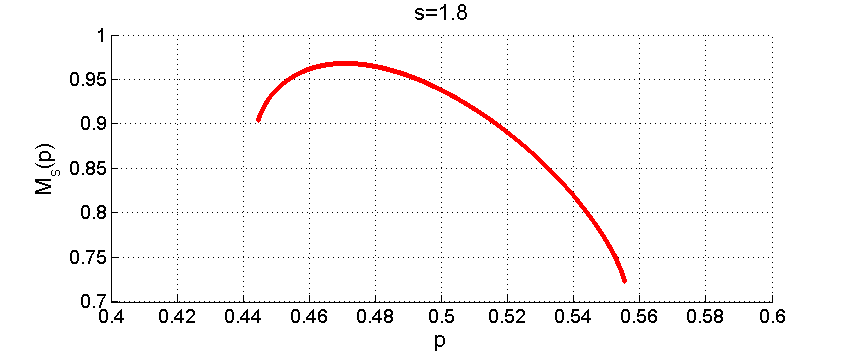}}{}&
\subf{\includegraphics[width=0.5\textwidth]{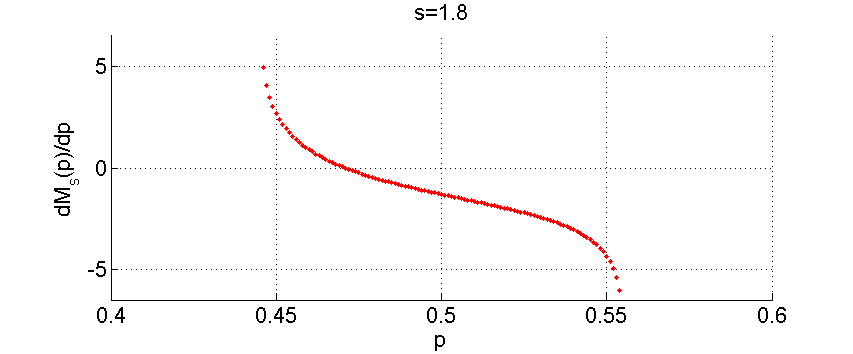}}{}
\\
\hline
        \end{tabular}
 \caption{The typical course of Information-Firing-Quotient $M_{s}$ versus spiking probability $p$ for larger jumping parameter $s>1$. Panels a, b course of $M_{s}$ and its derivative for $s=1.4$ is shown. Panels c, d the course of $M_{s}$ and its derivative for more jumping case i.e. $s=1.8$ is shown. Observe, that the graphs for $s>1$ are qualitatively different than for $s<1$, see Fig. \ref{fig1}.
}\label{fig2}
     \end{figure}
\subsection{Information Source of Bernoulli Type}
Let us consider the Information-Firing-Quotient formula for Information Source being a Bernoulli process with probability parameter $p$. For this source we denote the $IFQ$ by $B(p)$.
\par
Using (\ref{ITR_3}) and (\ref{ITR_2}) we get
\begin{equation}\label{Bp1}
B(p):=\frac{H_{1}(p)}{p} \ 
\end{equation}
for $0<p\le 1$.
Now, we evaluate the derivative of the quotient $B(p)$ over the firing rate $p$ (corresponding to energy), which has the form
\begin{equation}\label{dbdt}
\frac{\mbox{d}B}{\mbox{d}p}(p)=\frac{\mbox{d}}{\mbox{d}p}\left(\frac{ITR(p)}{f_{R}(p)}\right)=\frac{1}{p^{2}}\ln{(1-p)}
\end{equation}
for $0<p\le 1$.
In order to find lower and upper bounds of these expressions we make use of the following inequalities 
\begin{equation}\label{bound1}
1-\frac{1}{x} \le \ln{x} \le x-1 \ ,
\end{equation}
which hold for $0 \le x$ and the inequality
\begin{equation}\label{4ln}
4(1-x)x\ln{2} \le H_{1}(x) \le \ln{2} \ ,
\end{equation}
which is true for $0<x\le 1$.
\par
Applying (\ref{bound1}) and (\ref{4ln}) to (\ref{Bp1}) we obtain the following bounds for $B(p)$
\begin{equation}\label{4ln2}
4(1-p)\le B(p)\ln{2} \le \frac{\ln{2}}{p}
\end{equation}
and making use of (\ref{dbdt}) and (\ref{bound1}) we get bound of the derivative of $B(p)$
\begin{equation}
-\frac{1}{(1-p)p} \le \frac{\mbox{d}B}{\mbox{d}p}(p) \le -\frac{1}{p} \ .
\end{equation}
Introducing the following notation
\begin{eqnarray}\label{3B}
L_{0}(p):=4(1-p) \ ,\label{3B1}
\\
U_{0}(p):=\frac{1}{p} \ ,\label{3B2}
\\
L_{1}(p):=\frac{1}{(1-p)p} \ ,\label{3B3}
\end{eqnarray}
we express (\ref{4ln}) and (\ref{4ln2}) in the compact forms
\begin{eqnarray}
L_{0}(p) \ln{2}\le B(p) \le U_{0}(p) \ln{2} \ , \label{4ln3}
\\
-L_{1}(p) \le \frac{\mbox{d}B}{\mbox{d}p}(p) \le -U_{0}(p) \ . \label{4ln4}
\end{eqnarray}
Further on we show that these bounds can be interpreted as benchmarks for more complex processes such as Markov processes.
\subsection{Information Source of Markov Type}
Consider as an Information Source the two-state Markov process. Under the notation from Section \ref{markov_process} we introduce the "jumping" parameter $0 \le s \le 2$, which in fact can be interpreted as the tendency of transition from one state to the other state
\begin{equation}
s:=p_{0|1}+p_{1|0} \ .
\end{equation}
As we show below this parameter plays a crucial role in qualitative behavior of the $IFQ$ coefficient. Observe that for Markov case
\begin{equation}\label{p}
p=\frac{p_{1|0}}{p_{0|1}+p_{1|0}}=\frac{p_{1|0}}{s}
\end{equation}
\par
and for $0 \le s \le 1$ the firing frequency $p$ is in the full interval $[0,1]$, while for $1<s<2$ it is limited to the smaller interval $1-\frac{1}{s}<p<\frac{1}{s}$, i.e.
\begin{equation}
\frac{1}{2}-(\frac{1}{s}-\frac{1}{2}) \le p \le \frac{1}{2}+(\frac{1}{s}-\frac{1}{2}) \ .
\end{equation}
thus, $p$ is localized symmetrically around $\frac{1}{2}$. Note that for $s>1$ the spike probability $p$ is well separated from zero. 
\par
For the Markov source we denote the $IFQ$ indicator by $M$. Using (\ref{itr_markov_case}) and (\ref{p_markov_case}) we have
\begin{eqnarray}\label{m}
M:=\frac{ITR}{P_{eq}(1)}= 
\frac{P_{eq}(0)H_{1}(p_{1|0})+P_{eq}(1)H_{1}(p_{0|1})} {P_{eq}(1)} \ .
\end{eqnarray}
Making use of (\ref{p}) we express $M$ in terms of $p$ and $s$, 
\begin{equation}
M_{s}(p)=\frac{(1-p)H_{1}(ps)+pH_{1}((1-p)s)}{p}
\end{equation}
where $H_{1}$ is given by (\ref{H1}).
\par
Observe that for $s \le 1$ there are the following limits
\begin{eqnarray}
\lim_{p \to 0} {M_{s}}=+\infty \nonumber 
\quad \mbox{and} \quad \nonumber
\lim_{p \to 1} {M_{s}}=0 
\end{eqnarray}
and for $s>1$ 
\begin{equation}
\lim_{p \to 1-\frac{1}{s}} {M_{s}(p)}=\frac{[(1-s) \ln{(s-1)}-(2-s)\ln {(2-s)}]}{s-1}=\frac{H_{1}(s-1)}{s-1} \ ,
\end{equation}
\begin{equation}
\lim_{p \to \frac{1}{s}} {M_{s}(p)}=[(1-s)\ln{(s-1)}-(2-s)\ln{(2-s)}]=H_{1}(s-1) \ .
\end{equation}
Next we evaluate the derivative of the quotient $M_{s}(p)$ over the firing rate $p$
\begin{equation}\label{dMs}
\frac{\mbox{d}}{\mbox{d}p}M_{s}(p)=s\ln{(sp)} -s\ln{[1-(1-p)s]}-s\ln{(1-sp)}+s\ln{(s(1-p))}+\frac{\ln{(1-ps)}}{p^{2}}\ .
\end{equation}
Making use of (\ref{bound1}) and (\ref{4ln}) we obtain the following bounds
\begin{equation}
4(1-p)s(2-s)\ln{2} \le M_{s}(p) \le \frac{\ln{2}}{p}
\end{equation}
and referring to the Bernoulli bounds (\ref{3B1}) and (\ref{3B2}) we arrive to the following limits 
\begin{equation}\label{limits}
s(2-s)L_{0}(p)\ln{2} \le M_{s}(p)\le U_{0}(p)\ln{2} \ .
\end{equation}
Notice that the left bound is maximal for $s=1$, i.e. for Bernoulli case. 
\par
Now applying again (\ref{bound1}) and (\ref{4ln}) to (\ref{dMs}) we get
\begin{equation}
\frac{1}{s}\left[\frac{(s-1)s}{p(1-p)(1-ps)}-\frac{1}{p(1-p)}\right] \le \frac{\mbox{d}}{\mbox{d}p}M_{s}(p)\le -s\left[\frac{1}{p}-\frac{(s-1)}{[1-(1-p)s](1-sp)}\right]
\end{equation}
and referring again to the Bernoulli bounds (\ref{3B2}) and (\ref{3B3}) we obtain the following limits on derivative of $IFQ$ for Markov sources
\begin{equation} \label{last}
-\frac{1}{s}\left[L_{1}(p)-\frac{(s-1)s}{p(1-p)(1-ps)}\right] \le \frac{\mbox{d}}{\mbox{d}p}M_{s}(p) \le -s\left[U_{1}(p)-\frac{(s-1)}{[1-(1-p)s](1-sp)}\right] \ .
\end{equation}
Observe, that for $s=1$, inequalities (\ref{limits}) and (\ref{last}) reduce to inequalities (\ref{4ln3}) and (\ref{4ln4}) respectively, i.e. just to Bernoulli case. Moreover, we see that for $s$ close to 1 the bounds of $M_s(p)$ and $\frac{\mbox{d}}{\mbox{d}p}M_{s}(p)$ rigorously approach the Bernoulli bounds. For (\ref{limits}) this observation is clear while for (\ref{last}) it follows from the fact that the deviations from Bernoulli bounds, $L_{1}(p)$ and $U_{1}(p)$, contain the factor ($s$-1). 
\par
We see that for $0 \le s \le 1$ the derivative  is negative, $\frac{\mbox{d}}{\mbox{d}p}M_{s}(p)<0$, thus $M_{s}(p)$ is a decreasing function of $p$ (Fig. \ref{fig1}) and clearly it is significantly larger for small $p$. 
For $1<s<2$ case the behavior of function $M_{s}(p)$ is qualitatively different (Fig. \ref{fig2}). It is non-monotonic and it has a global maximum. What means that in this case for each s the optimal firing rate exists. 
\par
The increasing $s$ corresponds to the increasing $p_{1|0}$ so in this case the transition from the no-spike-state (state 0) to the spike-state (state 1) occurs more and more often. This means that for larger $s$ the neuron is more firing leading to bursting phenomena. 
\section{Discussion and Conclusions}
In this paper we address the fundamental question in neuronal coding. We analyze the possible correspondence between "temporal" and "firing rate" coding for two qualitatively different types of Information Sources. For the first type of source it is assumed that consecutive spikes are uncorrelated, thus it is governed by the Bernoulli process. In the second case we assume that there is a short time correlation (memory) between consecutive spikes, thus we model this source by the Markov process. 
\par
For the quantitative study of the relation between temporal and rate coding we propose the Information-Firing-Quotient being the ratio of Information Transmission Rate and firing rate. Since the energy used for transfer of information is proportional to the firing rate this quotient is understood as amount of information transmitted at the cost of unit energy. Clearly, for larger $IFQ$ the transmission is more efficient. The goal is to find the optimal parameters of transmission. We found that the crucial role in qualitative and quantitative behavior of $IFQ$ is played by the parameter $s$ which, in fact, measures the ability of transition from non-spike to spike state and vice versa. 
Taking into account that in the real biological systems the firing rate is limited from below by the spontaneous activity and very small values of $p$ and large values of $IFQ$, are non-realistic. This situation may happen for $s \le 1$ (Fig. \ref{fig1}), when $IFQ$ is monotonically decreasing with $p$, and then the realistic cut-off of $p$ separating it from zero should be assumed. On the other hand for $s>1$, i.e. for more active, say bursting, neurons we observe that the global maximum of $IFQ$ exists, thus in this case there is the unique optimal firing rate (Fig. \ref{fig2}) well separated from zero. This leads to the non-intuitive hypothesis that even for bursting phenomenon there may still exist the optimal regime of transmission. 
\section*{Acknowledgements}
We gratefully acknowledge financial support from the Polish National Science Centre under grant no. 2012/05/B/ST8/03010.
%
%
%
\section*{References}
%
%

%
\end{document}